\documentclass[aps,showpacs,prep,graphics,nofootinbib]{revtex4}%
\usepackage{amssymb}
\usepackage[dvips]{graphicx}
\usepackage{amsmath}
\usepackage{bm}
\usepackage{epsfig}
\usepackage{amsfonts}%
\providecommand{\U}[1]{\protect\rule{.1in}{.1in}}
\begin{document}
\title{Gauge invariant fluctuations of the metric during inflation from new
scalar-tensor Weyl-Integrable gravity model}
\author{$^{1}$ M. L. Pucheu,$^{1}$ C. Romero, $^{2,3}$ M. Bellini and $^{4}$ Jos\'e
Edgar Madriz Aguilar \thanks{E-mail address: madriz@mdp.edu.ar}. }
\affiliation{$^{1}$ Departamento de F\'{\i}sica, Universidade Federal da Para\'{\i}ba}
\affiliation{Caixa Postal 5008, 58051-970 Jo\~{a}o Pessoa, Pb, Brazil.}
\affiliation{E-mail: cromero@fisica.ufpb.br, mlaurapucheu@fisica.ufpb.br}
\affiliation{$^{2}$ Departamento de F\'isica, Facultad de Ciencias Exactas y Naturales,
Universidad Nacional de Mar del Plata (UNMdP), Funes 3350, C.P. 7600, Mar del
Plata, Argentina }
\affiliation{$^{3}$ Instituto de Investigaciones F\'{\i}sicas de Mar del Plata (IFIMAR)-
Consejo Nacional de Investigaciones Cient\'{\i}ficas y T\'ecnicas (CONICET) Argentina.}
\affiliation{E-mail: mbellini@mdp.edu.ar}
\affiliation{and }
\affiliation{$^{4}$ Departamento de Matem\'aticas, Centro Universitario de Ciencias Exactas
e Ingenier\'{\i}as (CUCEI), Universidad de Guadalajara (UdG), Av. Revoluci\'on
1500 S.R. 44430, Guadalajara, Jalisco, M\'exico. }
\affiliation{E-mail: madriz@mdp.edu.ar, edgar.madriz@red.cucei.udg.mx }

\begin{abstract}
We investigate gauge invariant scalar fluctuations of the metric during
inflation in a non-perturbative formalism in the framework of a recently
formulated scalar-tensor theory of gravity, in which the geometry of spacetime
is that of a Weyl integrable manifold. We show that in this scenario the Weyl
scalar field can play the role of the inflaton field. As an application of the
theory, we examine the case of a power law inflation. In this case, the
quasi-scale invariance of the spectrum for scalar fluctuations of the metric
is achieved for determined values of the parameter $\omega$ of the
scalar-tensor theory. We stress the fact that in our formalism the physical
inflaton field has a purely geometrical origin.

\end{abstract}

\maketitle

\vskip .5cm Weyl-Integrable geometry, FRW metric, inflation, scalar
fluctuations of the metric, scalar-tensor theory of gravity.

\section{Introduction}

The existence of an inflationary stage \cite{infl,infl2,infl3,infl4} of the early
universe is now supported by many observational evidences, in particular, by
the\ discovery of temperature anisotropies present in the cosmic microwave
background (CMB) \cite{re1,re1p}. In fact, in recent years there has been an
extraordinary development on observational tests of inflationary models
\cite{Bi2}. On the theoretical side, among the most popular and pioneering
models of inflation we would like to mention the supercooled chaotic
inflation\ model \cite{re2}. In this proposal, as we know, the expansion of
the universe is driven by a scalar field known as the inflaton field.\newline

In the current state of affairs one would say that an inflationary model is
considered viable when, among other features, it provides a mechanism for the
creation of primordial density fluctuations, needed to explain the subsequent
structure formation in the universe \cite{re3, mukhanov, wands}. According to some authors,
fluctuations of the inflaton field induce scalar fluctuations of the metric
around a Friedmann-Robertson-Walker (FRW) geometrical background. On the other
hand, scalar fluctuations of the metric on cosmological scales can be studied
in a non-perturbative formalism, describing not only small fluctuations, but
the larger ones \cite{re4}. In this kind of model, it is assumed that the
inflaton field exists at the beginning of the inflationary stage \cite{bcms}.
In view of this, it seems rather appealing and, perhaps, closer to the spirit
of general relativity, to look for a model that explains the origin of the
inflaton scalar field at a purely geometrical level.\newline

We all know that in the so-called scalar-tensor theories of gravity, in
addition to the space-time metric, a scalar field is required to describe
gravity. Historically, an early motivation for this new degree of freedom came
from the attempt to incorporate Mach's principle in a relativistic theory of
gravity \cite{re5,re6}. Later, new classes of scalar-tensor theories whose
main motivation has a geometric character have appeared \cite{geom}. Among
these, there has been an increasing interest in gravitational theories defined
in a Weyl integrable space-time geometry \cite{Novello}. These may be regarded
as scalar-tensor theories in which the scalar field plays a clear geometric
role \footnote{Incidentally, it should be noted that general relativity can
also be formulated in the language of Weyl ntegrable spacetime \cite{re6pp}.
In cosmology, it was shown that this new formulation of general relativity in
terms of the Weyl-Integrable geometry admits solutions capable to explain the
present accelerating expansion of the universe, as a natural consequence of
the existence of the Weyl scalar field \cite{re6ppp}. \newline}. \ In the
present paper, we shall consider a more recent geometrical approach to
scalar-tensor theory \cite{Pucheu}. This approach starts by considering the
action of Brans-Dicke theory and introduces the space-time geometry from first
principles, which is done by applying the Palatini formalism. Cosmological
models in this new geometrical scalar-tensor theory have been studied, and
cosmological scenarios have been found which seems to indicate the presence of
a geometric phase transition of the universe \cite{Pucheu2} .\

In this paper, we consider gauge invariant scalar fluctuations of the metric
during inflation employing a non-perturbative formalism in the context of the
Weyl scalar-tensor theory of gravity. Here, the physical inflaton field has a
geometrical origin since it is part of the affine structure of the space-time
manifold. As an application of this idea, we study a scenario in which the
early universe underwent an expansion phase given by power-law inflation. Our approach is different
than one earlier worked\cite{EPJC2009}, in which the inflaton field has a physical origin. In this
work we demonstrate that the expansion of the universe can be driven by a geometrical field (in a Weyl frame), which can be interpreted as the inflaton field when
we use the Einstein frame. The paper is organized as follows. In Section II, we present the general formalism
that leads to Weyl geometrical scalar-tensor theory. We proceed to Section
III, where we work out the formalism of gauge invariant scalar fluctuations of
our model in a non-perturbative way. The particular case of the dynamical equations for the small scalar fluctuations of the metric are studied using the
linearized case. The power spectrum and the mean square scalar fluctuations of the metric in a more general way are calculated at the end of this section.
In Sect. IV we present the example when the universe describes a power-law inflationary expansion. Finally, in section V, we conclude with some comments.

\section{The formalism}

To begin with, let us now consider a scalar-tensor theory of gravity whose
action is given by
\begin{equation}
S=\frac{1}{16\pi}\int d^{4}x\sqrt{-g}\left\{  \Phi R+\frac{\tilde{\omega}%
(\Phi)}{\Phi}g^{\mu\nu}\Phi_{,\mu}\Phi_{,\nu}-\tilde{V}(\Phi)\right\}, \label{i1}%
\end{equation}
where $R$ denotes the Ricci scalar, $\tilde{\omega}(\Phi)$ is a function of
the scalar field, $\tilde{V}(\Phi)$ is a scalar potential.

It is easy to see that in terms of the new variable $\phi=-ln(G\Phi)$, the
action (\ref{i1}) can be rewritten as
\begin{equation}
S=\int d^{4}x\sqrt{-g}\left\{  e^{-\phi}\left[  \frac{R}{16\pi G}+\omega
(\phi)g^{\mu\nu}\phi_{,\mu}\phi_{,\nu}\right]  -V(\phi)\right\}  \label{i2}%
\end{equation}
where we have defined $\omega(\phi)=(16\pi G)^{-1}\tilde{\omega}\left[
\Phi(\phi)\right]  $ and $V(\phi)=\tilde{V}\left(  \Phi(\phi)\right)  $.
Adopting the Palatini procedure, it can be shown that the variation of the
action (\ref{i2}) with respect to the affine connection yields \cite{Pucheu}
\begin{equation}
\nabla_{\alpha}g_{\mu\nu}=\phi_{,\alpha}g_{\mu\nu}, \label{i3}%
\end{equation}
which corresponds to the non-metricity condition characterizing a Weyl
integrable space-time, where $\phi$ is interpreted as the Weyl scalar field.
We thus see that if we adopt the Palatini variational principle we are
naturally led to the geometry of the Weyl integral space-time. To get the
complete set of field equations, we next perform the variation of the action
(\ref{i2}) with respect to the metric $g_{\alpha\beta}$ and the scalar field
$\phi$, which then gives
\begin{align}
G_{\mu\nu}  &  =8\pi G\left[  \omega(\phi)\left(  \frac{1}{2}g_{\mu\nu}%
\phi_{,\alpha}\phi^{,\alpha}-\phi_{,\mu}\phi_{,\nu}\right)  -\frac{1}%
{2}e^{\phi}g_{\mu\nu}V(\phi)\right] ,\label{i4}\\
\square\phi &  =-\left(  1+\frac{1}{2\omega(\phi)}\frac{d\omega(\phi)}{d\phi
}\right)  \phi_{,\mu}\phi^{,\mu}-\frac{e^{\phi}}{\omega(\phi)}\left(  \frac
{1}{2}\frac{dV}{d\phi}+V\right)  , \label{i5}%
\end{align}
where $G_{\mu\nu}=R_{\mu\nu}-(1/2)Rg_{\mu\nu}$ is calculated in terms of the
affine connection $\Gamma_{\mu\nu}^{\alpha}$, given by
\begin{equation}
\Gamma_{\mu\nu}^{\alpha}=\left\{  \,_{\mu\nu}^{\alpha}\right\}  -\frac{1}%
{2}g^{\alpha\beta}\left[  g_{\beta\mu}\phi_{,\nu}+g_{\beta\nu}\phi_{,\mu
}-g_{\mu\nu}\phi_{,\beta}\right]  , \label{cuw1}%
\end{equation}
$\left\{  \,_{\mu\nu}^{\alpha}\right\}  =(1/2)g^{\alpha\beta}(g_{\beta\mu,\nu
}+g_{\beta\nu,\mu}-g_{\mu\nu,\beta})$ denoting the Levi-Civita
connection.\newline

An important fact to be noted here is that the non-metricity condition
(\ref{i3}) is invariant under the Weyl transformations
\begin{align}
\overline{g}_{\mu\nu}  &  =e^{f}g_{\mu\nu},\label{inv1}\\
\overline{\phi}  &  =\phi+f, \label{inv2}%
\end{align}
where $f$ is an arbitrary scalar function of the coordinates. It is usually
said in the literature that the transformations (\ref{inv1})-(\ref{inv2}) lead
from one{ }\textit{frame} $(M,g,\phi)$ to another \textit{frame} $(M,\bar
{g},\bar{\phi})$. For the particular choice $f=-\phi$, we have $\bar{g}%
_{\mu\nu}=e^{-\phi}g_{\mu\nu}$ and $\bar{\phi}=0$, and in this case the
condition (\ref{i3}) reduces to the Riemannian metricity condition; and
because of this the frame $(M,\bar{g},\bar{\phi}=0)$ is referred to as
the \textit{Einstein frame.} However, the terminology \textit{Einstein frame} used in here is different from the traditional employed in Jordan-Brans-Dicke (JBD) scalar-tensor theories. This is because in the JBD theories the frame transformations do not preserve the compatibility of the metric and the affine connection, generating geodesics in the traidtional Einstein frame with an extra acceleration term. In here, the \textit{Riemann} or \textit{Einstein frame} is a geometric object constructed as a result of the invariance under Weyl transformations of the non-metricity condition (\ref{i3}), and thus the geodesics are preserved in all Weyl frames, including the now called \textit{Einstein frame}, which is defined by means of the effective metric $\bar{g}_{\mu\nu}$.  Now, it is not difficult to verify that in the
Einstein frame the action (\ref{i2}) takes the form
\begin{equation}
S^{(R)}=\int d^{4}x\sqrt{-\overline{g}}\left\{  \frac{\overline{R}}{16\pi
G}+\omega(\phi)\overline{g}^{\mu\nu}\phi_{,\mu}\phi_{,\nu}-e^{2\phi}%
V(\phi)\right\} , \label{i6}%
\end{equation}
with $\overline{R}$ denoting the transformed Riemannian Ricci scalar. The
field equations derived from the action (\ref{i6}) will be given by
\begin{align}
&  \overline{G}_{\mu\nu}=8\pi G\left[  \omega(\phi)\left(  \frac{1}{2}%
\phi_{,\alpha}\phi^{,\alpha}\overline{g}_{\mu\nu}-\phi_{,\mu}\phi_{,\nu
}\right)  -\frac{e^{2\phi}}{2}\overline{g}_{\mu\nu}V(\phi)\right]  ,\label{i7}\\
&  \overline{\square}\phi=-\frac{1}{2\omega}\frac{d\omega}{d\phi}\phi
_{,\alpha}\phi^{,\alpha}-\frac{e^{2\phi}}{\omega}\left(  V+\frac{1}{2}%
\frac{dV}{d\phi}\right)  , \label{i8}%
\end{align}
where $\overline{G}_{\mu\nu}$ and $\overline{\square}$ denote the Einstein
tensor and the D'Alembertian operator, respectively, both calculated with the
affine connection in the Einstein frame \footnote{Here the energy
momentum-tensor $T_{\mu\nu}$ is defined according the prescription adopted in
\cite{Pucheu}.}.\newline

In order to study scalar fluctuations of the metric during inflation, we shall
consider the simplest scalar-tensor theory derived from (\ref{i7}) and
(\ref{i8}). This is achieved when we choose the parameter $\omega(\phi)$ to be
a constant. In this case, the field equations (\ref{i7})-(\ref{i8}) reduce to
\begin{align}
&  \overline{G}_{\mu\nu}=8\pi G\left[  \omega\left(  \frac{1}{2}\phi_{,\alpha
}\phi^{,\alpha}\overline{g}_{\mu\nu}-\phi_{,\mu}\phi_{,\nu}\right)
-\frac{e^{2\phi}}{2}\overline{g}_{\mu\nu}V(\phi)\right] ,\label{i10}\\
&  \overline{\square}\phi=-\frac{e^{2\phi}}{\omega}\left(  V+\frac{1}{2}%
\frac{dV}{d\phi}\right). \label{i11}
\end{align}
One interesting feature of this framework is that, in contrast to what happens
in a general Weyl frame, in the Einstein frame the scalar field $\phi$ is no
longer a geometric field, and should be regarded as a physical field. In this
way, we shall consider $\phi$ in (\ref{i10})-(\ref{i11}) as the inflaton field,
the scalar field that drives the expansion of the universe during inflation.
From this point of view the term between brackets in the right side of (\ref{i10}) can be
interpreted as an induced energy-momentum tensor in the Einstein frame
\begin{equation}
\overline{T}_{\mu\nu} = \omega\left(  \frac{1}{2}\phi_{,\alpha
}\phi^{,\alpha}\overline{g}_{\mu\nu}-\phi_{,\mu}\phi_{,\nu}\right)
-\frac{e^{2\phi}}{2}\overline{g}_{\mu\nu}V(\phi).
\end{equation}
For convenience, we shall work in the Einstein frame, although the results hold
in a general frame \footnote{It is important to note that in Weyl geometrical
scalar-tensor theory all geometrical objects, such as geodesics, curvature and
length of a curve, are constructed in a invariant way with respect to Weyl
transformations. This is done with the help of the "effective" metric
$\gamma_{\mu\nu}=e^{-\phi}g_{\mu\nu}$, which is a fundamental invariant of the
equivalence class of Weyl manifolds. It turns out that in the Einstein frame
$\gamma_{\mu\nu}=\overline{g}_{\mu\nu}$. See, for instance, \cite{Pucheu}.}.

\section{Non-perturbative gauge invariant scalar fluctuations of the metric}

In order to study gauge invariant scalar fluctuations of the metric, let us just by generality, start by following the non-perturbative formalism introduced in \cite{re4}. Thus, the perturbed line element can be written in the form
\begin{equation}
ds^{2}=e^{2\psi}dt^{2}-a^{2}(t)e^{-2\psi}(dx^{2}+dy^{2}+dz^{2}), \label{b1}%
\end{equation}
where $a(t)$ is the cosmological scale factor and $\psi(t,x,y,z)$ is a metric
function describing gauge invariant scalar fluctuations of the metric in a
non-perturbative way. The Ricci scalar calculated with the metric in
(\ref{b1}) is given by
\begin{equation}
\bar{R}=6e^{-2\psi}\left[  \frac{\ddot{a}}{a}+H^{2}-\ddot{\psi}-5H\dot{\psi
}+3\dot{\psi}^{2}+\frac{e^{4\psi}}{3a^{2}}(\nabla^{2}\psi-(\nabla\psi
)^{2})\right]  , \label{b2}%
\end{equation}
with $H=\dot{a}/a$ denoting the Hubble parameter. In the absence of matter,
the perturbed field equations (\ref{i10}) reduce to
\begin{align}
e^{-2\psi}(3H^{2}-6H\dot{\psi}+3\dot{\psi}^{2})+\frac{1}{a^{2}}\left[
2\nabla^{2}\psi-(\nabla\psi)^{2}\right]  e^{2\psi}  &  =8\pi G\left[
\frac{\omega}{2}\left(  e^{-2\psi}\dot{\phi}^{2}+\frac{e^{2\psi}}{a^{2}%
}(\nabla\phi)^{2}\right)  +\frac{1}{2}e^{2\phi}V(\phi)\right]  ,\label{b6}\\
(-2\ddot{\psi}+5\dot{\psi}^{2}-8H\dot{\psi}+\frac{2\ddot{a}}{a}+H^{2}%
)e^{-2\psi}-\frac{1}{3a^{2}}(\nabla\psi)^{2}e^{2\psi}  &  =8\pi G\left[
-\frac{\omega}{2}\left(  e^{-2\psi}\dot{\phi}^{2}-\frac{1}{3a^{2}}e^{2\psi
}(\nabla\phi)^{2}\right)  +\frac{1}{2}e^{2\phi}V(\phi)\right]  ,\label{b3}\\
\frac{1}{a}\frac{\partial}{\partial x^{i}}\left[  \frac{\partial}{\partial
t}(a\psi)\right]  -\frac{\partial\psi}{\partial t}\frac{\partial\psi}{\partial
x^{i}}  &  =8\pi G\frac{\omega}{2}\frac{\partial\phi}{\partial t}%
\frac{\partial\phi}{\partial x^{i}},\label{b4}\\
\frac{\partial\psi}{\partial x^{i}}\frac{\partial\psi}{\partial x^{j}}  &
=-8\pi G\frac{\omega}{2}\frac{\partial\phi}{\partial x^{i}}\frac{\partial\phi
}{\partial x^{j}},\label{b5}
\end{align}
while the dynamics of the inflaton field $\phi$ is given by
\begin{equation}
\ddot{\phi}+(3H-4\dot{\psi})\dot{\phi}-\frac{e^{4\psi}}{a^{2}}\nabla^{2}%
\phi+\frac{1}{\omega}e^{2(\psi+\phi)}\left[  V(\phi)+\frac{1}{2}V^{\prime
}(\phi)\right]  =0, \label{b7}%
\end{equation}
where the prime mark denotes derivative with respect to $\phi$. It is not
difficult to verify that after some algebraic manipulations the equations
(\ref{b6}) and (\ref{b3}) yield
\begin{equation}
(\ddot{\psi}-4\dot{\psi}^{2}+7H\dot{\psi})e^{-2\psi}+\frac{2}{3a^{2}}%
(\nabla\psi)^{2}e^{2\psi}-\frac{1}{a^{2}}(\nabla^{2}\psi)e^{2\psi}=8\pi
G\left[  -\frac{\omega}{3a^{2}}e^{2\psi}(\nabla\phi)^{2}-\frac{1}{2}e^{2\phi
}V(\phi)\right]  , \label{b8}%
\end{equation}
where we have eliminated background contributions. This equation determines
the dynamics of $\psi$, the function that describes the scalar fluctuations of
the metric with arbitrary amplitude \footnote{We would like to point out that
the non-perturbative method adopted here is an extension of the standard
approach. The main difference lies in the fact that in the non-perturbative
approach cosmological curvature fluctuations of the metric with large
amplitude can be also treated, whereas in the standard approach the
fluctuations must be small compared to the metric background. A more complete
description of the non-perturbative method can be found in the reference [7].}.

\subsection{The linear approximation}

In general, finding solutions for the dynamics of $\psi$ \ from (\ref{b8}) is
not an easy task. However, we can obtain some solutions in the weak field
limit, i.e., solutions corresponding to small amplitudes of the scalar
fluctuations. In this limit, a linear approximation of the scalar fluctuations
of the metric is useful and the gauge invariance is preserved. In order to
implement this limit, we use $e^{\pm n\psi(x^{\sigma})}\simeq1\pm
n\psi(x^{\sigma})$. In this regime, a semiclassical approximation for the
inflaton field is also valid. Thus we write $\phi(t,x^{i})=\phi_{b}%
(t)+\delta\phi(t,x^{i})$, where the background classical field $\phi
_{b}=\left\langle E|\phi|E\right\rangle $ is the expectation value of $\phi$,
with $\left.  |E\right\rangle $ denoting a physical quantum state given by the
Bunch-Davies vacuum \cite{BD}, and $\delta\phi$ describes the quantum
fluctuations of the field $\phi$. The line element (\ref{b1}) in the weak
field limit becomes
\begin{equation}
ds^{2}=(1+2\psi)dt^{2}-a^{2}(t)(1-2\psi)(dx^{2}+dy^{2}+dz^{2}). \label{b9}%
\end{equation}
Here, the metric $\overline{g}_{\mu\nu}^{(0)}=\mathrm{diag}\left[
1,-a^{2}(t),-a^{2}(t),-a^{2}(t)\right]  $ describes the background metric in
the Einstein frame, which is supposed to be isotropic and homogenous. Notice
that the linear approximation (\ref{b9}) agrees with the longitudinal gauge in
the standard approach to perturbation theory \cite{bardeen,riotto}. The
linearization of the equation (\ref{b8}) leads to
\begin{equation}
\ddot{\psi}+7H\dot{\psi}-\frac{1}{a^{2}}\nabla^{2}\psi=-4\pi Ge^{2\phi_{b}%
}V^{\prime}(\phi_{b})\delta\phi. \label{b10}%
\end{equation}
From the equations (\ref{b4}) and (\ref{b5}), (\ref{b10}) can be put in the
form
\begin{equation}
\ddot{\psi}+\alpha(t)\dot{\psi}-\frac{1}{a^{2}}\nabla^{2}\psi+\beta(t)\psi=0, \label{equation01}
\end{equation}
where
\begin{equation}
\alpha(t)=7H+\frac{e^{2\phi_{b}}}{\omega\dot{\phi}_{b}}V^{\prime}(\phi_{b}),
\label{equation10}%
\end{equation}%
\begin{equation}
\beta(t)=\frac{e^{2\phi_{b}}}{\omega}\left(  V(\phi_{b})+\frac{V^{\prime}%
(\phi_{b})}{\dot{\phi_{b}}}H\right)  . \label{equation11}%
\end{equation}
On the other hand, with respect to the background part (i.e. on cosmological
scales) the linearization of the equation (\ref{b7}) gives
\begin{equation}
\ddot{\phi}_{b}+3H_{c}\dot{\phi}_{b}+\frac{e^{2\phi_{b}}}{\omega}\left[
V(\phi_{b})+\frac{1}{2}V^{\prime}(\phi_{b})\right]  =0, \label{b12}%
\end{equation}
whereas, on small quantum scales, the dynamics of $\delta\phi$ and $\psi$ is
given by
\begin{equation}
\ddot{\delta\phi}+3H_{c}\dot{\delta\phi}-\frac{1}{a^{2}}\nabla^{2}\delta
\phi+\frac{e^{2\phi_{b}}}{\omega}\left[  V^{\prime}(\phi_{b})+\frac{1}%
{2}V^{\prime\prime}(\phi_{b})\right]  \delta\phi=4\dot{\phi}_{b}\dot{\psi
}-\frac{2e^{2\phi_{b}}}{\omega}\left[  V(\phi_{b})+\frac{1}{2}V^{\prime}%
(\phi_{b})\right]  \psi. \label{b13}%
\end{equation}
At the same time, the Friedmann equation for the background metric is
\begin{equation}
3H_{c}^{2}=4\pi G\left[  \omega\dot{\phi}_{b}^{2}+e^{2\phi_{b}}V(\phi
_{b})\right]  . \label{b14}%
\end{equation}
Finally, from (\ref{b12}) and (\ref{b14}) we find that the background inflaton
field satisfies the equation
\begin{equation}
\dot{\phi}_{b}^{2}=-\frac{\dot{H}_{c}}{4\pi G\omega}. \label{b15}%
\end{equation}
Now, in order to study the quantum dynamics of the inflaton field $\phi$ and
of the scalar fluctuations of the metric $\psi$, we shall follow the
cannonical quantization procedure.

\subsection{Quantization and spectrum for scalar fluctuations of the metric}

Following the cannonical quantization procedure of quantum field theory, we start by imposing the commutation relation
\begin{equation}
\lbrack\psi(\overline{x}),\Pi_{(\psi)}^{0}(t,\overline{x}^{\prime})\rbrack=i \delta^{(3)}(\overline{x}-\overline{x}^{\prime}),  \label{equation02}%
\end{equation}
where the quantity $\Pi_{(\psi)}^{0}=\partial L/\partial\dot{\psi}$ is the
cannonical conjugate momentum to $\psi$ and $L$ is the Lagrangian. The equal times quantization condition (\ref{equation02}) implies that
\begin{equation}
\lbrack\psi(t,\overline{x}),\dot{\psi}(t,\overline{x}^{\prime})]=\frac
{i}{t_{0}}\,e^{-\int\alpha(t)\,dt}\delta^{(3)}\left(  \vec{x}-\vec{x}^{\prime
}\right)  . \label{qcond}%
\end{equation}
According to the action (\ref{i6}), now $L$ takes the form
\begin{equation}
L=\sqrt{-\overline{g}}\left[  \frac{\bar{R}}{8\pi G}-\omega\overline{g}%
^{\mu\nu}\phi_{,\mu}\phi_{,\nu}-e^{2\phi}V(\phi)\right]  , \label{equation03}%
\end{equation}
where $\bar{R}$ is given by the expression (\ref{b2}).\newline

In order to simplify the structure of the equation (\ref{equation01}), we can introduce the auxiliary field
\begin{equation}
\psi(\bar{x},t)=e^{-\frac{1}{2}\int\alpha(t)dt}\chi(\bar{x},t), \label{eeqq1}%
\end{equation}
so that (\ref{equation01}) can be written in terms of $\chi$
\begin{equation}
\ddot{\chi}-\frac{1}{a^{2}}\nabla^{2}\chi+\left[  \beta-\left(  \frac
{\alpha^{2}}{4}+\frac{\dot{\alpha}}{2}\right)  \right]  \chi=0. \label{eeqq2}%
\end{equation}
The auxiliary field $\chi(\bar{x},t)$ can be expanded in Fourier modes as
\begin{equation}
\chi(\bar{x},t)=\frac{1}{(2\pi)^{3/2}}\int d^{3}k\left[  a_{k}e^{i\bar{k}%
\cdot\bar{x}}\xi_{k}(t)+a_{k}^{\dagger}e^{-i\bar{k}\cdot\bar{x}}\xi_{k}^{\ast
}(t)\right]  \label{eeqq3}%
\end{equation}
with the asterisk mark denoting complex conjugate, $a_{k}^{\dagger}$ and
$a_{k}$ denoting the creation and annihilation operators, which satisfy the
commutation relations
\begin{equation}
\left[  a_{k},a_{k^{\prime}}^{\dagger}\right]  =\delta^{(3)}(\vec{k}-\vec
{k}^{\prime}),\quad\left[  a_{k},a_{k^{\prime}}\right]  =\left[
a_{k}^{\dagger},a_{k^{\prime}}^{\dagger}\right]  =0. \label{eeqq4}%
\end{equation}
From (\ref{eeqq3}) we can see that the equation (\ref{eeqq2}) leads to
\begin{equation}
\ddot{\xi}_{k}+\left[  \frac{k^{2}}{a^{2}}-\left(  \frac{\alpha^{2}}{4}%
+\frac{\dot{\alpha}}{2}-\beta\right)  \right]  \xi_{k}=0, \label{eeqq5}%
\end{equation}
which determines the dynamics of the quantum modes $\xi_{k}$ for the scalar
fluctuations of the metric. The squared quantum fluctuations of $\psi$ in the
IR-sector (cosmological scales) are given by the expression
\begin{equation}
\left\langle \psi^{2}\right\rangle _{IR}=\frac{e^{-\int\alpha(t)dt}}{2\pi^{2}%
}\int_{0}^{\epsilon k_{H}}\frac{dk}{k}k^{3}\left.  \left[  \xi_{k}(t)\xi
_{k}^{\ast}(t)\right]  \right\vert _{IR}, \label{eeqq6}%
\end{equation}
where $\epsilon=k_{max}^{IR}/k_{p}\ll1$ is a dimensionless parameter,
$k_{max}^{IR}=k_{H}(t_{r})$ \ being the wave number related to the Hubble
radius at time $t_{r}$, when the modes re-enter the horizon, while $k_{p}$ is
the Planckian wave number. For the Hubble parameter $H=0.5\times10^{-9}M_{p}$,
the values of $\epsilon$ range between $10^{-5}$ and $10^{-8}$, and this
corresponds to a number of e-foldings at the end of inflation $N_{e}=63$.

\section{An example: A power law inflation}

As an application of the ideas developed previously, we now obtain the
spectrum of squared scalar quantum fluctuations of the metric $\psi$ in the
framework of \ the Weyl geometric scalar-tensor theory in the case of \ a
power law inflationary expansion of the universe.\newline

Considering the power law expansion $a(t)=a_{0}(t/t_{0})^{p}$, the equation
(\ref{b15}) leads to the classical solution
\begin{equation}
\phi_{b}(t)=\phi_0 \left[1+ n_0\,ln\left(  \frac{t}{t_{e}}\right) \right], \label{b16}
\end{equation}
where $\phi_{0}=\phi_{b}(t_{0})$, $t_{0}$ being the time when inflation starts. For the potential, if we substitute (\ref{b15})
in (\ref{b14}), we then get
\begin{equation}\label{VV}
e^{2\phi_{b}/\phi_0} V(t)=\frac{3H_{c}^{2}+\dot{H}_{c}}{4\pi G},
\end{equation}
which, for $H_{c}=p/t$, reduces to
\begin{equation}
V(t)=\frac{1}{4\pi G}\frac{p(3p-1)}{t_{e}^{-2n_{0}}}t^{-2(1+n_{0})},
\label{pex2}%
\end{equation}
where
\begin{equation}
n_{0}=-\sqrt{p/(4\pi G\phi_{0}^{2}\omega)}, \label{n0}
\end{equation}
is a negative dimensionless parameter. Taking into account the equations (\ref{equation10}) and
(\ref{equation11}), the expression (\ref{eeqq5}) can be written in the form
\begin{equation}
\ddot{\xi}_{k}+\left(  \frac{\kappa^{2}}{t^{2p}}-\frac{\gamma_{0}^{2}}{t^{2}%
}\right)  \xi_{k}=0, \label{pex3}%
\end{equation}
being $\gamma_{0}^{2}=\alpha_{0}^{2}/4-\alpha_{0}/2-\beta_{0}$ and $\kappa
^{2}=(k^{2}t_{0}^{2p})/a_{0}^{2}$. The parameters $\alpha_{0}$ and $\beta_{0}$
are given by
\begin{equation}
\alpha_{0}=7p-\frac{2p(3p-1)(1+n_{0})}{4\pi G\omega n_{0}^{2}},\quad\beta
_{0}=\frac{p(3p-1)}{4\pi G\omega}\left[  1-\frac{2(1+n_{0})}{n_{0}^{2}%
}e^{2\phi_{0}}\right]  . \label{pex4}%
\end{equation}
The general solution of (\ref{pex3}) reads
\begin{equation}
\xi_{k}(t)=A_{1}\sqrt{t}\,\mathcal{H}_{\nu}^{(1)}[Z(t)]+A_{2}\sqrt
{t}\,\mathcal{H}_{\nu}^{(2)}[Z(t)], \label{pex5}%
\end{equation}
with $\nu=\sqrt{4\gamma_{0}^{2}+1}/(2p-2)$, $Z(t)=\kappa t^{1-p}/(p-1)$,
$A_{1}$ and $A_{2}$ being integration constants. The functions $\mathcal{H}%
_{\nu}^{(1,2)}$ denote the first and second kind Hankel functions,
respectively. The expression (\ref{qcond}) in terms of the $\chi$ field
becomes
\begin{equation}
\left[  \chi(\bar{x},t),\dot{\chi}(\bar{x}^{\prime},t)\right]  =\frac{i}%
{t_{0}}\delta^{(3)}\left(  \bar{x}-\bar{x}^{\prime}\right)  , \label{pex6}%
\end{equation}
and thus the normalization condition for the modes will be given by
\begin{equation}
\dot{\xi}_{k}^{\ast}\xi_{k}-\xi_{k}^{\ast}\dot{\xi}_{k}=\frac{i}{t_{0}}.
\label{pex7}%
\end{equation}
From (\ref{pex7}) and by choosing the Bunch-Davies condition, the normalized
solution for the modes $\xi_{k}$ is
\begin{equation}
\xi_{k}(t)=i\sqrt{\frac{\pi}{4t_{0}}}\left(  \frac{1}{p-1}\right)
^{\frac{1-p}{2}}\sqrt{t}\,\mathcal{H}_{\nu}^{(2)}[Z(t)]. \label{pex8}%
\end{equation}
The mean square fluctuations for $\psi$ on the IR sector according to the
equation (\ref{eeqq6}) are then given by
\begin{equation}
\left\langle \psi^{2}\right\rangle _{IR}=\frac{(-2a_{0})^{2\nu}}{8\pi^{3}%
}\frac{\Gamma^{2}(\nu)}{(p-1)^{1-p-2\nu}}\frac{t^{1-2\nu(1-p)-\alpha_{0}}%
}{t_{0}^{1-2p(3+\nu)}}\frac{\epsilon^{3-2\nu}}{3-2\nu}\beta_{0}^{2}%
t_{r}^{2(p-1)(3-2\nu)}, \label{pex9}%
\end{equation}
where we have used that $k_{H}=(\gamma_{0}^{2}/t_{0}^{2p})t_{r}^{2(p-1)}$. The
corresponding power spectrum for $\psi$ has the form
\begin{equation}
\mathcal{P}_{k}(\psi)=\frac{(-2a_{0})^{2\nu}}{8\pi^{3}(3-2\nu)}\frac
{\Gamma^{2}(\nu)}{(p-1)^{1-p-2\nu}}\frac{t^{1-2\nu(1-p)-\alpha_{0}}}{t^{1+2\nu
p}}k^{3-2\nu}. \label{pex10}%
\end{equation}
%The quasi-scale invariance is achieved when $\nu\simeq3/2$. This condition
%holds when the expression
%\begin{equation}
%9(p-1)^{2}\simeq(7p-\frac{d_{0}}{\omega})^{2}-14p+\frac{d_{0}-d_{1}}{\omega
%}+1, \label{pex11}%
%\end{equation}
%is satisfied. The constant parameters $d_{0}$ and $d_{1}$ are given by
%\begin{equation}
%d_{0}=\frac{p(3p-1)(1+n_{0})}{2\pi Gn_{0}^{2}},\quad d_{1}=\frac{p(3p-1)}{\pi
%G}\left[  1-\frac{2(1+n_{0})}{n_{0}^{2}}e^{2\phi_{0}}\right]  . \label{pex12}%
%\end{equation}
%Solving (\ref{pex11}) for $\omega$ we obtain
%\begin{equation}
%\omega\simeq\frac{\pm(d_{1}+14d_{0}p-d_{0})+d_{2}}{8(10p^{2}+p-2)},
%\label{pex13}%
%\end{equation}
%where
%\begin{equation}
%d_{2}=\sqrt{d_{1}^{2}+28d_{1}d_{0}p-2d_{1}d_{0}+36d_{0}^{2}p^{2}-44d_{0}%
%^{2}p+33d_{0}^{2}}. \label{pex14}%
%\end{equation}
We then see that the quasi scale invariance for the spectrum of scalar
fluctuations of the metric $\mathcal{P}_{k}(\psi)$ is achieved for the values
of $P/\rho \gtrsim -1$. The spectral index is given
by $n_{s}-1=3-2\nu$, so that once $n_{s}$ is known, \ we can determine the
parameter $p$ of the power law expansion of the universe, which then will be
given by
\begin{equation}
p=1+\frac{2}{1-n_{s}}.
\end{equation}
For $n_{s}\simeq0.96$\cite{RPP1}, we obtain $p\simeq 51$. For this spectral
index, the equation of state $P=-\left(  \frac{2\dot{H}}{3H^{2}}+1\right)
\,\rho$ will lead to
\begin{equation}
P/\rho=\frac{2-3p}{3p}\simeq-0.9869,
\end{equation}
where $P$ is the pressure, and $\rho$, the energy density. This is in good
agreement with the data obtained by the WMAP9-$\omega$CDM(flat) observations
\cite{RPP2}. This is very interesting because, since $P/\rho=-\frac{p^{2}}{4\pi G\phi_{0}^{2}}$, for $n_0=-1$ we obtain from (\ref{b16}) that $\phi_0$ is
\begin{equation}
\phi_0 \simeq \frac{3.57}{\sqrt{\pi \omega}}\,G^{-1/2},
\end{equation}
where we have taken into account the value $p=51$. Hence, this means that $\phi_0$ takes sub Planckian values for $\omega> 4.06$, which solves the problem of the trans Planckian values in models of standard inflation. Furthermore, the choice $n_0 =-1$ corresponds to a constant $V(t)$ and $\gamma^2_0=-\beta_0=\frac{p(1-3p)}{4\pi G\omega}$. With this choice for $n_0$, and taking into account the expression (\ref{VV}), is possible to set the following correspondence with standard models of inflation:
\begin{equation}
{\cal V}(\phi_b) = 2 V(t) e^{2\phi_b/\phi_0}.
\end{equation}
In other words, the effective potential ${\cal V}(\phi_b)$ can be interpreted potential in standard models of inflation for $n_0=-1$. This is very interesting because now it is possible
to define the slow roll parameters of inflation evaluated at $k=a H$: $\epsilon= \frac{1}{16\pi G} \left(\frac{{\cal V}'}{{\cal V}}\right)^2$, $\eta = \frac{1}{8\pi G} \frac{{\cal V}"}{{\cal V}}$. In our case we obtain that
\begin{equation}
2\epsilon=\eta= \frac{1}{2\pi G \phi_0^2}= 0.03887.
\end{equation}
With this value, one can automatically calculate the scalar and tensor spectral indices: $n_s=1-6\epsilon+2\eta=1-\eta=0.9612$, $n_t\simeq -2\epsilon=-\eta=-0.0387$. Therefore, is
immediate the calculation of the ratio between both indices: $r=n_t/n_s = -\eta/(1-\eta)=-0.04$. This value is in agreement with observations\cite{RPP1}.

\section{ Final Comments}

In this paper, we have considered gauge invariant scalar fluctuations of the
cosmological metric during the inflationary phase of the universe in the
framework of the Weyl geometrical scalar-tensor theory of gravity and using a
non-perturbative formalism. \ By investigating the limit of the field
equations in the case of small perturbations we have obtained the spectrum of
scalar fluctuations at the end of inflation. An important feature of this
model is that the inflaton field is modeled by a geometrical scalar field,
which is part of the affine structure of the background geometry. As far as
inflationary models are concerned, it is perhaps tempting to take the view
that a geometrical origin of the inflaton might be regarded as more natural
than its introduction a priori without a clear theoretical justification.

\section*{Acknowledgements}

\noindent  The authors would like to thank CONACYT and Guadalajara University (M\'{e}xico), UNMdP and CONICET (Argentina), CAPES and CNPQ (Brazil) for financial support.


\bigskip

\begin{thebibliography}{99}                                                                                               %


\bibitem {infl}A. A. Starobinsky, Phys. Lett. \textbf{B91}: (1980) 99;

\bibitem {infl2}A. H. Guth, Phys. Rev. \textbf{D23}: (1981) 347;

\bibitem {infl3}A. D. Linde, Phys. Lett. \textbf{B129}: (1983) 177.

\bibitem{infl4}V. Mukhanov, \textit{Physical Foundations of Cosmology}, Cambridge University Press, Cambridge, (2005). A. R. Liddle and D. H. Lyth, \textit{Cosmological Inflation and Large-Scale Structure},  Cambridge University Press, (2000). A. D. Linde, \textit{Particle Physics and Inflationary Cosmology}, Contemporary Concepts in Physics Vol. 5, (1990).

\bibitem {re1}N. Aghanim, {\em et al}, (Planck Collaboration) (2015), ArXiv:1507.02704.

\bibitem {re1p}K. A. Olive, et al (Particle Data Group) Chin. Phys. \textbf{C38} (2014) 090001.

\bibitem {Bi2}P. A. R. Ade, \emph{et al}, Phys. Rev. Lett. \textbf{114} (2015) 101301.

\bibitem {re2}A. Linde, Phys. Lett. \textbf{B116}, (1982) 335.

\bibitem{mukhanov} V.F. Mukhanov, H.A. Feldman and R.H. Brandenberger, Phys. Rept. \textbf{215} (1992) 203.

\bibitem{wands}  K. A. Malik and D. Wands, Phys. Rept. \textbf{475} (2009) 1.

\bibitem {re3}M. Anabitarte, M. Bellini, J.E. Madriz-Aguilar, Eur. Phys. J. \textbf{C65} (2010) 295.

\bibitem {re4}M. Anabitarte, M. Bellini, Eur. Phys. J. \textbf{C60} (2009) 297.

\bibitem {bcms}M. Bellini, H. Casini, R. Montemayor, P. Sisterna, Phys. Rev. \textbf{D54} (1996) 7172.

\bibitem {re5}C. H. Brans, R. H. Dicke, Phys. Rev. \textbf{124} (1961) 925.

\bibitem {re6}V. Faraoni, Annals Phys. \textbf{317} (2005) 366-382.
%\bibitem{re6p} I. Quiros, R. Garc\'{i}a-Salcedo, J. E. Madriz-Aguilar, T. Matos, Gen. Rel. Grav. {\bf 45} (2013) 489-518.


\bibitem {geom} See, for instance, P. C. Peters, J. Math. Phys. {\bf 10} (1969) 1029; \\
G. Lyra, Math. Z. {\bf 54} (1951) 52;\\
H. H. Soleng, Class. Quantum Grav. {\bf 5} (1988) 1489;\\ 
D. K. Sen, Z. Phys. {\bf 149} (1957) 311; \\ 
D. K. Ross, Gen. Rel. Grav. {\bf 6} (1975) 157; \\ 
K. A. Dunn, J. Math. Phys. {\bf 15} (1974) 2229; \\ 
J.B. Fonseca-Neto, C. Romero, S.P.G. Martinez, Gen .Rel. Grav. \textbf{45} (2013) 1579.

\bibitem {Novello} M. Novello, L. A. R. Oliveira, J. M. Salim, E. Elbas, Int. J.
Mod. Phys. \textbf{D1} (1993) 641;\\ 
J. M. Salim and S. L. Saut\'{u}, Class. Quant. Grav. \textbf{13} (1996) 353; \\ 
H. P. de Oliveira, J. M. Salim and S. L. Saut\'{u}, Class. Quant. Grav. \textbf{14} (1997) 2833; \\ 
V. Melnikov, \textit{Classical Solutions in Multidimensional Cosmology} in Proceedings of the VIII Brazilian School of Cosmology and Gravitation II
(1995), edited by M. Novello (Editions Fronti\`{e}res) pp. 542-560, ISBN 2-86332-192-7; \\ 
K.A. Bronnikov, M.Yu. Konstantinov, V.N. Melnikov, Grav. Cosmol. {\bf 1} (1995) 60; \\ 
J. Miritzis, Class. Quantum. Grav. \textbf{21} (2004) 3043; \\
J. Miritzis, J. Phys. Conf. Ser. \textbf{8} (2005) 131; \\
J. E. M. Aguilar and C. Romero, Found. Phys.\textbf{\ 39} (2009) 1205; \\
J. E. M. Aguilar and C. Romero, Int. J. Mod. Phys. {\bf A24} (2009) 1505; \\
J. Miritzis, Int. J. Mod. Phys. \textbf{D22} (2013) 1350019; \\ 
F. P. Poulis and J. M. Salim arXiv:1305.6830; \\ 
R. Vazirian, M. R. Tanhayi and Z. A. Motahar, Adv. High Energy Phys. \textbf{7} (2015) 902396; \\ 
I. P. Lobo, A. B. Barreto, and C. Romero, Eur. Phys. J. {\bf C75} (2015) 448.

\bibitem {Pucheu}T. S. Almeida, M. L. Pucheu, C. Romero, J. B. Formiga, Phys.
Rev. \textbf{D89} (2014) 064047.

\bibitem {Pucheu2}M.L. Pucheu, F.A.P. Alves-Junior, A.B. Barreto, C. Romero,
e-Print: arXiv:1602.06966.

\bibitem{EPJC2009} M. Anabitarte, M. Bellini, Eur. Phys. J. {\bf C60} (2009) 297.

\bibitem {re6pp}C. Romero, J. B. Fonseca-Neto, M. L. Pucheu, Class. Quant.
Grav. \textbf{29} (2012) 155015.

\bibitem {re6ppp}R. Aguila, J. E. Madriz-Aguilar, C. Moreno, M. Bellini, Eur. Phys. J. {C74} (2014) 3158.

\bibitem {BD}T. S. Bunch and P. C. W. Davies, Proc. Roy. Soc. \textbf{A360} (1978) 117.

\bibitem {bardeen}J. M. Bardeen, Phys. Rev. \textbf{D22} (1980) 1882.

\bibitem {riotto}D. H. Lyth, A. Riotto, Phys. Rept. \textbf{314} (1999) 1.

\bibitem {RPP1} O. Lahov, A. R. Liddle, Chin. Phys. \textbf{C38} (2014) 345.

\bibitem {RPP2}M. J. Mortonson, D. H. Weinbert, M. White, Chin. Phys. \textbf{C38} (2014) 361.
\end{thebibliography}
\end{document}